# Amino acid composition and thermal stability of protein structures: the free energy geography of the Protein Data Bank

**Antonio Deiana[1]\*, Kana Shimizu[2,3]\*, Andrea Giansanti[1,4]\*§**

[1]Department of Physics, Sapienza, University of Rome, P.le A. Moro, 5, 00185, Rome, Italy
[2]Department of Computer Science, Graduate School of Science and Engineering, Waseda University, 3-4-1 Okubo, Shinijuku-ku, Tokyo, 169-8555, Japan
[3]Computational Biology Research Center (CBRC), National Institute of Advanced Industrial Science and Technology (AIST), 135-0064, 2-4-7 Aomi, Koto-ku, Tokyo, Japan
[4]INFN, Sezione di Roma, P.le A. Moro, 5, 00185, Rome, Italy

\*These authors contributed equally to this work
§Corresponding author

Email addresses:
    AD: Antonio.Deiana@roma1.infn.it
    KS: shimizu-kana@aist.go.jp
    AG: Andrea.Giansanti@roma1.infn.it

# Abstract

We study the combined influence of amino acid composition and chain length on the thermal stability of protein structures. A new parameterization of the internal free energy is considered, as the sum of hydrophobic effect, hydrogen-bond and de-hydration energy terms. We divided a non-redundant selection of protein structures from the Protein Data Bank into three groups: i) rich in order-promoting residues (OPR proteins); ii) rich in disorder-promoting residues (DPR proteins); iii) belonging to a twilight zone (TZ proteins). We observe a partition of PDB in several groups with different internal free energies, amino acid compositions and protein lengths. Internal free energy of 96% of the proteins analyzed ranges from -2 to -6.5 kJ/mol/res. We found many DPR and OPR proteins with the same relative thermal stability. Only OPR proteins with internal energy between -4 and -6.5 kJ/mol/res are observed to have chains longer than 200 residues, with a high de-hydration energy compensated by the hydrophobic effect. DPR and TZ proteins are shorter than 200 residues and they have an internal energy above -4 kJ/mol/res, with a few exceptions among TZ proteins. Hydrogen-bonds play an important role in the stabilization of these DPR folds, often higher than contact energy. The new parameterization of internal free energy let emerge a geography of thermal stabilities of PDB structures. Amino acid composition per se is not sufficient to determine the stability of protein folds, since. DPR and TZ proteins generally have a relatively high internal free energy, and they are stabilized by hydrogen-bonds. Long DPR proteins are not observed in the PDB, because their low hydrophobicity cannot compensate the high de-hydration energy necessary to accommodate residues within a highly packed globular fold.

# Background

Protein folding is the assembly process that drives a natively synthesized protein to get a compact, equilibrium three-dimensional structure. The stability of protein native structure has been for long time strictly connected with functional activity [1, 2]. There is consensus that the hydrophobic effect is the main determinant of the folding process, since it gives the highest negative free energy contribution that compensates for the loss of conformational entropy, implied in the compaction of the structure [3-6].

A bias in amino acid composition towards hydrophobicity or hydrophilicity, therefore, might play an important role in protein folding. In the last decade it has been widely recognized that many proteins rich in hydrophilic residues are natively unfolded, i.e. they lack a well-defined tertiary structure [7-9]. Folded and natively unfolded proteins are differently biased in their amino acid compositions, the latter being enriched in R, K, E, P, and S (disorder-promoting residues, mainly hydrophilic) and depleted in C, W, Y, I, and V (order-promoting residues, mainly hydrophobic) [10]. It should be also noted that amino acid composition is the most important pattern considered by methods that predict whether a protein sequence folds into a tertiary structure or not [11, 12].

From a thermodynamic point of view, protein folding can be viewed as a pathway in the free energy configuration landscape, towards a ground-state conformation [13, 14]. Folded and unfolded proteins are supposed to have quite different free energy landscapes, those of natively unfolded proteins being characterized by many minima separated by low energy barriers [15]. Amino acid composition might significantly affect thermodynamic stability of proteins by modulating the ruggedness of the free energy landscape.

Several papers, on the other hand, question the above scenario. In particular it has been proposed that, besides amino acid composition, hydrogen-bonds give a relevant contribution to the energetics of a protein fold [16, 17].

A sharp view is that of Ghosh and Dill, who argue that amino acid composition does not affect in a systematic way thermodynamic stability, which is solely determined by the length of the protein chain. Using calorimetry data collected by Robertson and Murphy [18], they observe that many folded proteins have universal values of enthalpy and entropy per residue. Based on this observation, they introduce a model to estimate the free energy of unfolding of protein sequences. The model should be valid for many folded proteins, denoted by Ghosh and Dill as thermally ideal [19].

The synthetic review of the literature given above is sufficient to state that there is still an open problem about the relevance of amino acid composition for the thermodynamic stability of protein folds and the aim of this work is then to make a phenomenological survey of the stability of PDB structures, using a simple but significant form of statistical potential. We considered a set of generic, not ligated, non-redundant and not disordered structures of the Protein Data Bank (PDB) [20] and then, using our previously proposed $S_{SU}$ classifier as a sieve [21], we partitioned the dataset into three groups, characterized by different amino acid compositions: i) structures rich in order-promoting residues (OPR structures); ii) rich in disorder-promoting residues (DPR structures); iii) structures with about the same percentage of order- and disorder-promoting residues, belonging to a twilight zone (TZ proteins) [21, 22]. The relative thermal stability of the structures within these groups was estimated using a coarse-grained internal free energy $E_{INT}$, which is the sum of two terms related to contact energy and to hydrogen-bond energy (see equations (3) and (4) in methods). If amino acid composition were the only determinant of thermal stability then, expectedly, DPR proteins (more hydrophilic) would be systematically less stable than OPR ones (more hydrophobic). At variance with this expectation the panorama is more articulated. There are relevant energy and length thresholds, implied by the observation that structures with more than 200 residues seem to require OPR composition and an internal energy below -4.2 kJ/mol/res to get a standard stable fold.

The problem of correlating amino acid composition with thermal stability of the structure is then more subtle than expected. Amino acid composition modulates the values of the internal free energy per residue $e_{INT}$, but also the ratio between contact and hydrogen-bond energy, and, more in detail, also the contribution of the hydrophobic effect to the contact energy. A dominant contribution of contact energy to $e_{INT}$ is shown to be a necessary condition for long proteins to get a stable fold. Interestingly, in proteins shorter than 200 residues OPR or DPR compositions are not strictly correlated with different thermal stabilities. In particular, richness in disorder-promoting residues is only a sufficient condition for a short structure to become less stable than the standard based on the Ghosh-Dill (GD) model [19]. The ratio between contact and hydrogen-bond energy seems to be the real relevant parameter, which measures the propensity of a protein sequence to get a hydrophobic core and to get a relevant entropy loss upon folding.

It is interesting to point out that three well-performing predictors of disorder, combined in $S_{SU}$, erroneously predict about 5% of the proteins in our dataset as natively unfolded, despite the fact that their structure is fully solved. The $e_{INT}$ of these anomalously classified structures is systematically above that of more hydrophobic proteins, correctly classified as folded, and is often dominated by hydrogen-bonding. Moreover, a group of poorly stable, short DPR proteins is characterized by the weakest $e_{INT}$ among PDB structures, corresponding to open, non- globular structures.

In the following section we present our results, in detail. First, we show that the partition operated by $S_{SU}$ is, in fact, a separation in hydrophobicity. Then the energetics of the dataset is investigated, and, based on their $e_{INT}$, a geography of PDB structures is proposed. We discuss the contributions to $e_{INT}$ due to contact energy, hydrogen-bonding, hydrophobic effect and de-hydration energy and their possible relation with amino acid composition and protein length. The peculiar case of short, hyperstable structures stabilized by disulphide-bridges is also discussed in detail.

# Results

**OPR and DPR proteins are separated in the charge – hydrophobicity plane.**

To study whether amino acid composition is related to protein internal free energy or not, we selected 616 protein structures from the PDB, not ligated, non-homologous and without disordered residues (see methods). The consensus index SSU [21] was able to separate the 616 initially selected structures into: i) 395 (64%) OPR structures; ii) 30 (5%) DPR structures; iii) 191 (31%) structures with about the same percentage of order- and disorder-promoting residues, belonging to a twilight zone in the amino acid composition space (defined above as TZ proteins) [21, 22].

The three predictors combined in $S_{SU}$ (see Methods) classified OPR and DPR structures as folded and unfolded, respectively. OPR and DPR structures are well separated in the charge-hydrophobicity (CH) plane (supplementary figure 1). Their mean normalized hydrophobicity is 0.467 ± 0.001 and 0.401 ± 0.008, respectively. Since DPR proteins are hydrophilic, in accordance with the criterion by Uversky et al. [7, 9], they would be classified as natively unfolded.

Consistently with their definition, TZ structures largely overlap both with OPR and DPR proteins, and a significant number of them have an intermediate hydrophobicity (see inset of supplementary figure 1), belonging to a twilight area in the CH plane. Mean normalized hydrophobicity of TZ proteins is 0.434 ± 0.002, intermediate between OPR and DPR proteins. Incidentally, the observation that at least 5% of the dataset might be considered as consensus false predictions of good predictors, sets an upper bound to specificity. Assuming, as usual, the rate of false predictions to be 1-Sp, Sp does not exceed then 0.95, which should be considered, at present, as the standard performance of sequence only, composition based, predictors of natively unfolded proteins, consistent with previous reports [23].

We point out that in this work $S_{SU}$ has been used as a mere separator of protein structures into three families with amino acid compositions which are, respectively, similar to those of folded, natively unfolded and twilight zone proteins and therefore rich in order-promoting residues (OPR), disorder-promoting residues (DPR) and belonging to the twilight zone (TZ). The focus is not on natively unfolded proteins, but on the energetics of PDB structures.

**Geography of thermal stability of the PDB**

The internal free energies per residue $e_{INT}$ (see equations (2) – (4) in Methods) was evaluated in the different families, and plotted as a function of protein length N (figure 1).

In figure 1 red and blue dots represent, respectively, OPR and DPR proteins while black dots represent proteins from the reference dataset of Robertson and Murphy [18]. The free energy of folding of these thermally ideal proteins scales linearly with chain length and is assumed independent from amino acid composition and structural characteristics [19]. We note however that these thermally ideal proteins are mainly OPR. The $e_{INT}$ of these proteins falls between -2.0 and -6.5 kJ/mol/res. This range of energies, which comprises the majority of structures (96%) in the dataset, is delimited by two solid lines, traced to exactly include the black dots. In this way we delimit the internal free energies per residue of standard stable PDB structures, which are both OPR and DPR, short and long.

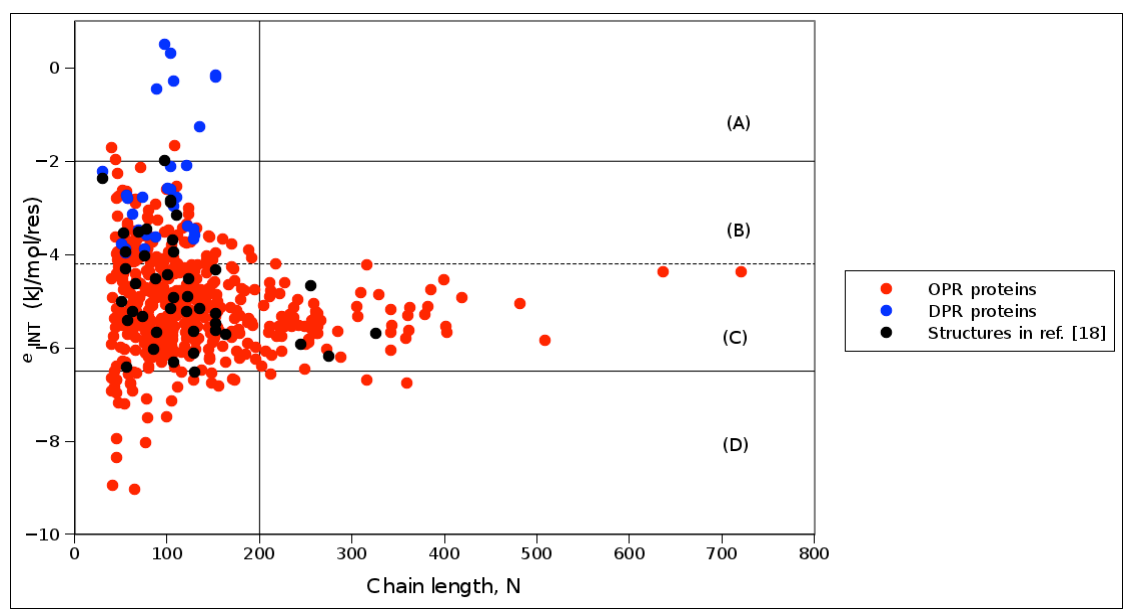

**Figure 1. Internal free energy per residue of OPR and DPR proteins in the PDB, as a function of their length N**
Structures in the PDB can be divided into four classes. Group A, DPR structures with internal free energy $e_{INT}$ above -2 kJ/mol/res. Group B, short OPR and DPR structures (less than 200 residues) with $e_{INT}$ between -2.0 and -4.2 kJ/mol/res. Group C, OPR proteins with $e_{INT}$ between -4.2 and -6.5 kJ/mol/res, with both short and long chains. Group D, short OPR structures with $e_{INT}$ below -6.5 kJ/mol/res. The two solid horizontal lines at -2.0 and -6.2 kJ/mol/res have been traced to delimit the range of internal energies typical of structures from the dataset of ref. [18] (black dots), assumed as a standard in the GD model [19]. The broken line at -4.2 kJ/mol/res was drawn to comprise in group C long chains present in the dataset. The vertical line divides short and long chains.

From the data in figure 1 we propose a classification of PDB structures into four groups, based on their energetics.
**Group A** contains 7 (1%) short (< 200 residues), NMR solved structures with $e_{INT}$ above -2 kJ/mol/res.
**Group B** contains short chains, 80 (13%) OPR and 23 (4%) DPR structures, with $e_{INT}$ between -2 and -4.2 kJ/mol/res.
**Group C** contains 228 (37%) short OPR and 63 (10%) long OPR structures with $e_{INT}$ between -4.2 and -6.5 kJ/mol/res.

**Group D** contains short hyper-stable structures with a peculiar amino acid composition, with an internal free energy below -6.5 kJ/mol/res.

As regards TZ proteins, they fall mainly into groups A and B (see supplementary figure 2).

Let us remind that the GD model predicts a free energy of folding "linearly dependent on chain length N, but with slope near 0 " [19]. It is then interesting to draw a graph of internal free energy as a function of chain length N in the different groups of structures, as shown in figure 2. The different groups of structures display different slopes, which correspond to the different values of $<E_{INT}>$ (see table 1 below). We propose here, phenomenologically, to distinguish three regions of the $E_{INT}$-N plane.

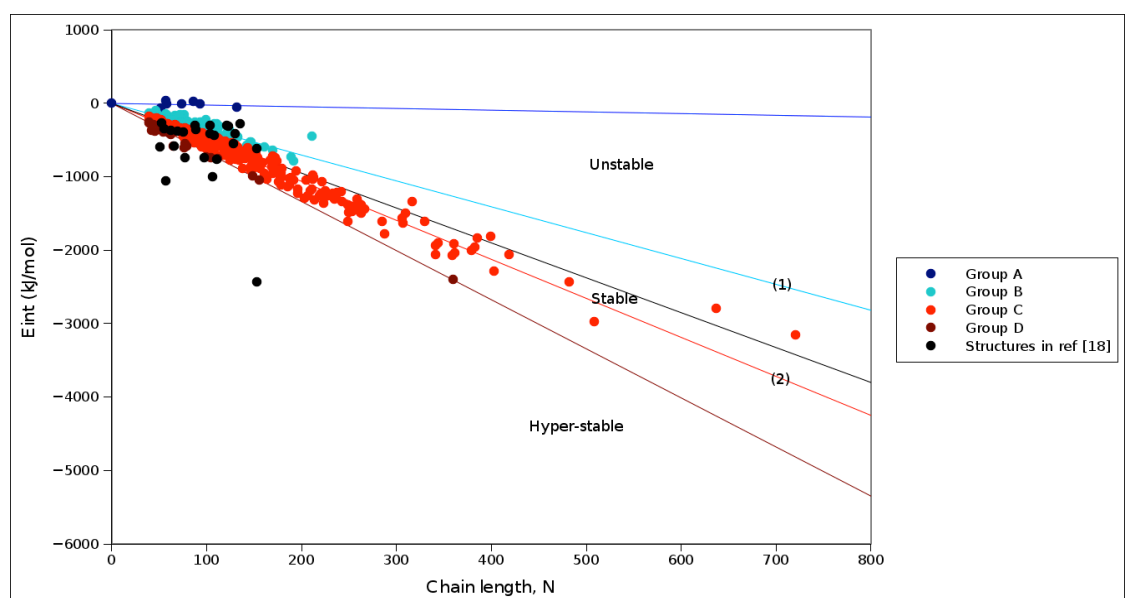

**Figure 2 Internal free energy as a function of chain length N**
The internal (not normalized) free energy is plotted as a function of the number of residues accommodated in the PDB structures belonging to the different groups introduced above (see figure 1). The solid black straight line with slope near 0 corresponds to the marginal stability of the GD model [19]. The other lines, with different slopes correspond to the different values of $<E_C>$ (per residue), in each group. The line associated with proteins in the list of Robertson and Murphy (black dots in figure 1) lies in the central part of the sector associated with thermally stable structures (see text) and constitutes the standard of thermal stability of the GD model.

The sector between lines 1 and 2 comprises structures in groups B and C, thermally stable. The sector above line 1 comprises thermally unstable structures and the sector below line 2 thermally hyper-stable structures. These terms could be misleading and, in order to avoid confusion, we specify that all structures are indeed marginally stable [2]. Stable, unstable and hyper-stable are here terms relative to the standard $-e_{INT}^{(REF)}$, estimated following the GD model [19] as -5.14 kJ/mol/res (see Methods eq.(10)) and introduce a phenomenological classification of protein structures

**Contact energy and hydroge-bond energy differently contribute to the internal free energy of PDB structures**

It is interesting to split $e_{INT}$ into the sum of contact energy $e_C$ [24] and hydrogen-bond energy $e_{HB}$ [25] (see eqs. (3) and (4) in methods) in the different groups of PDB structures classified above, following their energetics (Table 1).

| Structural group | $<e_{INT}>$ (kJ/mol/res) | $<e_C>$ (kJ/mol/res) | $<e_{HB}>$ (kJ/mol/res) | $<e_C/e_{HB}>$ |
|---|---|---|---|---|
| A | -0.2 ± 0.2 | 0.9 ± 0.2 | -1.1 ± 0.2 | -0.9 ± 0.2 |
| B, short DPR | -3.1 ± 0.1 | -1.4 ± 0.1 | -1.7 ± 0.1 | 0.9 ± 0.1 |
| B, short OPR | -3.62 ± 0.06 | -1.93 ± 0.07 | -1.69 ± 0.05 | 1.32 ± 0.09 |
| C, short OPR | -5.28 ± 0.04 | -3.29 ± 0.04 | -2.00 ± 0.03 | 1.76 ± 0.04 |
| C, long OPR | -5.40 ± 0.06 | -3.23 ± 0.06 | -2.17 ± 0.02 | 1.49 ± 0.03 |
| D | -7.3 ± 0.1 | -5.3 ± 0.2 | -2.0 ± 0.1 | 3.0 ± 0.3 |

**Table 1. Contributions of contact energy and hydrogen-bonds to the internal free energy of different groups of structures in the PDB.**
The different groups are defined by $<e_{INT}>$, the average internal free energy per residue, that is the sum of $<e_C>$ and $<e_{HB}>$ (see equations (3) and (4) in Methods). $<e_C>$ is the average contact energy per residue, estimated through the 1999 M-J potential; $<e_{HB}>$ is the hydrogen-bond energy per residue, estimated through DSSP. Short means less than 200 residues long, uncertainties are estimated as standard deviations from the average values.

In the proteins of group A hydrogen-bonding energy prevails over contact energy, which is anomalously positive.
In group B, DPR proteins are characterized by $<e_C/e_{HB}>$ ratios generally lower than 1, pointing to a greater contribution of hydrogen-bonding to the internal energy. On the contrary, OPR structures of group B have $<e_C/e_{HB}>$ ratios generally greater than 1.
Proteins in group C are all OPR structures, with $e_{INT}$ between -4.2 and -6.5 kJ/mol/res. In this group we find both short and long proteins, and their folds are as stable as the reference standard of thermally ideal proteins, with internal free energies of both groups comparable with the standard $e_{INT}^{(REF)}$ (-5.14 kJ/mol/res as reminded above). The average $<e_C/e_{HB}>$ ratio in these standard structures is close to 1.65.
It is remarkable that TZ structures have an average $<e_C/e_{HB}>$ ratio less than 1 when they belong to the unstable groups A and B, and greater than 1 in groups C and D (supplementary table 1). This indicates that $<e_C/e_{HB}>$ is strictly connected with relative stability, independently from amino acid composition.
Finally, proteins in group D are characterized by a quite low $e_{INT}$, below -6.5 kJ/mol/res. The structures in this group are rich in glycines and cysteines and, reasonably, their stability is mainly due to the formation of disulphide bridges as shown in figure 5 and discussed below. Since the internal free energy in this group is definitely below $e_{INT}^{(REF)}$ we call these structures hyper-stable, more stable than the standard, and with an $e_C$ on the average three times higher than $e_{HB}$.

**Contact energy results from the compensation between hydrophobic effect and de-hydration (de-mixing) energy**

Interestingly, in the short structures of group B, passing from the DPR to the OPR composition there is an increase of stability measured by a change in $e_{INT}$ of about 0.5 kJ/mol/res, which is totally reflected in an equal change in contact energy, the hydrogen-bonding term being insensitive, as expected, to amino acid composition. It is also interesting to observe that, passing from the short OPR structures of group B to the more stable short OPR structures in group C, the variation of $e_{INT}$ is 1.66 kJ/mol/res and that about 80% of this change is due to contact energy and the rest to hydrogen-bonding. Understanding the subtle balance of energy terms, due to composition but also to other still unclear factors in these three groups of particularly interesting short structures, deserves a more detailed investigation.

Following the paper of Li *et al.* [26] we split then contact energy into two contributions: the hydrophobic effect $e_{HP}$, estimated through Kauzmann's model [5], and a de-hydration (or de-mixing) term $\Delta e_{HYD}$, due to the breaking of residue-solvent interaction and the forming of residue-residue internal contacts (table 2). Incidentally, the de-hydration energy difference, as pointed out by Honig [3], should be destabilizing (i. e. positive), and this is what we observe. A quick inspection of the table 2 points out that the 0.5 kJ/mol/res change in $e_C$ between DPR and OPR structures of group B is due to a stabilization in hydrophobic effect energy by about 1kJ/mol/res and a de-stabilization by about 0.5 kJ/mol/res. Then, we conclude that, among the structures of group B, enrichment in order promoting residues implies a stabilization due to the hydrophobic effect, partly compensated by a destabilization due to the de-hydration term.

The change in contact energy of about 2 kJ/mol/res between the short OPR structures in group B with the more stable short OPR structures belonging to group C is totally due to a change in hydrophobic effect, but in this case the change in the hydrophobic effect evidently cannot be addressed to a change in the amino acid composition, since both groups of proteins are rich in OPR. Since the energetics of the hydrophobic effect in Kauzmann's model is due to the internalization of exposed surface (eq. (11) in methods) the change of about 2 kJ/mol/res in this term could be due to a compaction of the structure implying an evident increase in the number of contacts.

Looking then at the energy changes due to chain length in group C, it is evident that passing from short to long proteins implies a stabilizing change in hydrophobic effect, almost exactly compensated by a destabilization due to the change in de-hydration energy. Since the packing of protein structures increases with chain length and saturates above N=150 residues [27] we suggest that in long OPR structures of group C there is a change of $e_{HP}$ with respect to shorter structures, due to the increase in packing, which is compensated by 2kJ/mol/res of change in $\Delta e_{HYD}$ due to the increased difficulty for external residues to break contacts with the solvent to make internal contacts in a crowded interior saturated of contacts.

| Structural groups | $<e_C>$ (kJ/mol/res) | $<e_{HP}>$ (kJ/mol/res) | $<\Delta e_{HYD}>$ (kJ/mol/res) |
| --- | --- | --- | --- |
| A | 0.9 ± 0.2 | -6.8 ± 0.9 | 7.6 ± 1.0 |
| B, short DPR | -1.4 ± 0.1 | -6.9 ± 0.3 | 5.5 ± 0.3 |
| B, short OPR | -1.93 ± 0.07 | -7.9 ± 0.2 | 6.0 ± 0.2 |
| C, short OPR | -3.29 ± 0.04 | -9.3 ± 0.1 | 6.0 ± 0.1 |
| C, long OPR | -3.23 ± 0.06 | -11.26 ± 0.07 | 8.0 ± 0.1 |
| D | -5.3 ± 0.2 | -8.3 ± 0.4 | 3.0 ± 0.5 |

**Table 2. Splitting contact energy of PDB structures into hydrophobic effect and de-hydration energy**
The hydrophobic effect energy $e_{HP}$ is estimated through Kauzmann's model. The de-hydration energy $\Delta e_{HYD}$ is the gain in internal energy when two residue break a contact with the solvent and make an internal contact between them. For the sake of comparison contact energies $e_C$ are also reported.

To further elaborate on this compensation mechanism it is interesting to graph de-hydration energy as a function of contact energy (see figure 3). If we exclude long OPR structures of group C, $e_C$ and $\Delta e_{HYD}$ are linearly correlated (Pearson's correlation coefficient r = 0.90) whereas the correlation is significantly reduced if we include these proteins in the statistics (r = 0.63). This indicates that long OPR structures are outliers in the $\Delta e_{HYD} - e_C$ plane, having anomalously high values of $\Delta e_{HYD}$, possibly due to the high number of residues that must be accommodated within the interior of the globular fold.

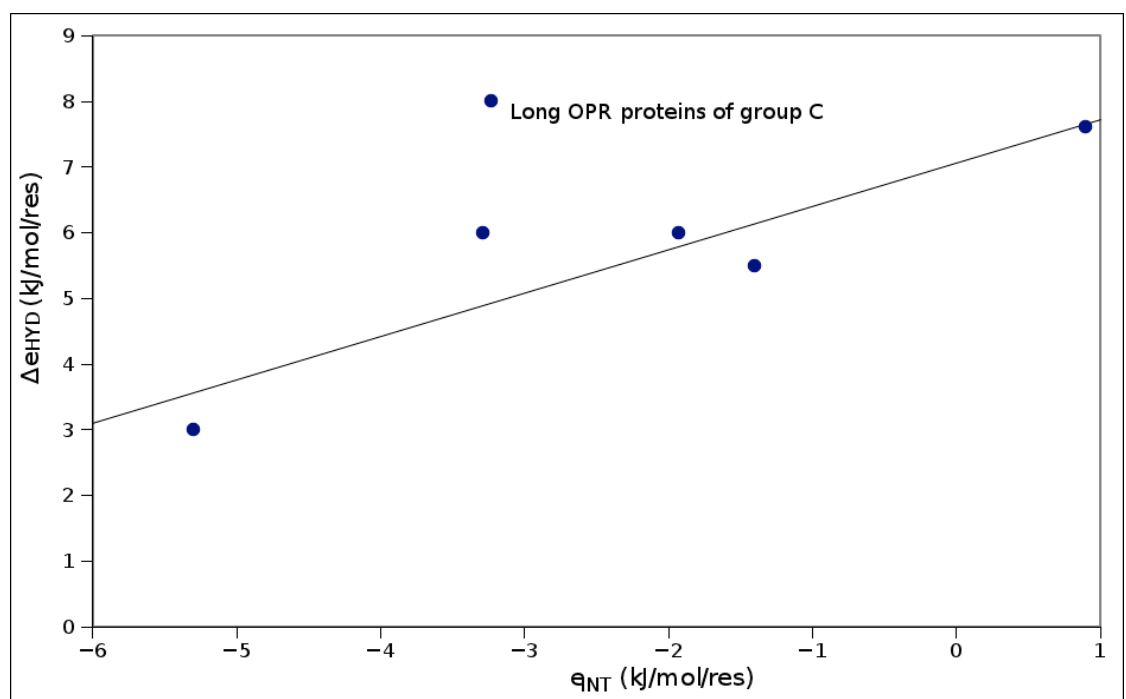

**Figure 3. De-hydration energy per residue plotted against internal free energy per residue.**
High values of the de-hydration energy correspond to high values of the internal free energy and therefore to low stability of protein folds. With the exception of long OPR proteins in group C, the correlation between de-hydration energy and internal free energy is almost linear.

Summarizing, unstable PDB structures are represented by a few DPR, NMR solved structures with an internal free energy above -2 kJ/mol/res and $e_C/e_{HB}$ ratios below 1, therefore stabilized mainly by hydrogen-bonds. Stable protein structures, are both OPR and DPR structures and they can have both short and long polypeptide chains. They have an internal energy ranging from -2 to -6.5 kJ/mol/res and $e_C/e_{HB}$ ratios between 1 and 2. Hyper-stable structures are characterized by $e_{INT}$ below -6.5 kJ/mol/res and with a contact energy that is at least twice the hydrogen bonding energy. Group C structures are mainly OPR, with just a few very short TZ peptides, and only within the range of internal free energies typical of this group is possible to find long PDB structures. Long proteins are characterized by the highest de-hydration energy compensated by hydrophobic effect. They are absent in A, B, and D groups. From the point of view of composition, to be rich in order-promoting residues seems then to be an essential prerequisite for a protein structure to accommodate more than 200 residues. From the point of view of energetics, an $e_{INT}$ comprised between -4.2 and -6.5 kJ/mol/res and, as shown in the next section, dominated by contact energy, is determinant to have a fully solved long protein structure. Amino acid composition rich in DPR seems sufficient to have an internal energy above -4.2 kJ/mol/res, although not necessary, since we also observe OPR proteins with this energy value. Conversely, richness in OPR seems a necessary condition for a protein to have an energy below -4.2 kJ/mol/res.

We checked that most part of TZ proteins belong to groups A and B (see supplementary figure 2). It is worth noting that very short TZ proteins (e.g. peptides with less than 40 residues) are present in all other groups, indicating that, particularly in the case of very short chains, a balance between order- and disorder-promoting residues makes the structure prone to have a wide range of internal free energies and, correspondingly of thermal stability. This observation suggests that amino acid composition affects energetics only in proteins with more than 40 residues.

**Unstable and hyper-stable structures**

To conclude this section let us focus a little more on unstable and hyper-stable structures. Group A proteins have a remarkably high destabilizing (positive) contact energy. In these proteins, the ratio $<e_C/e_{HB}>$ is significantly lower than 1 and negative, indicating that they are stabilized mainly by hydrogen-bonds. Their hydrophobic effect energy $e_{HP}$ is similar to that of DPR proteins of group B. However, group A proteins have a remarkably high de-hydration energy $\Delta e_{HYD}$ which is not compensated by the hydrophobic effect. This observation points out that in these proteins the energetic cost of burying amino acids in the protein interior not compensated by hydrophobic gain of stability could explain why these structures are quite open (see supplementary figure 3) and much less stable than the standard.

Group D proteins have a quite low contact energy, however their hydrogen-bonds and hydrophobic terms are similar to those in OPR proteins. We observe, in group D, a low de-hydration energy. Miyazawa and Jernigan state in their paper that contacts among cysteine residues give the highest contribution to contact energy in their potential [24]. Based on this observation, we verified that group D proteins are enriched in cysteine residues, moreover, all cysteine residues in our sample are involved in disulphide bridges. It is then reasonable to think that group D proteins are those short proteins, generally rich both in order- and disorder-promoting residues, that stabilize their tertiary structure through the formation of disulphide bridges (see figure 4).

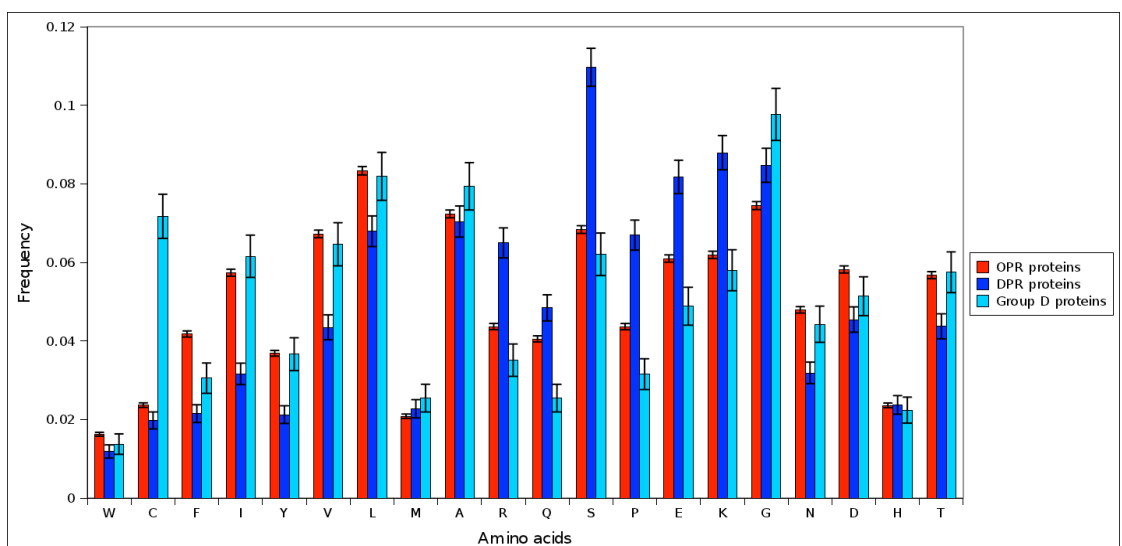

**Figure 4. Frequency of amino acids in OPR, DPR and in structures of group D**
Red bars refer to proteins rich in order-promoting residues (OPR), blue bars refer to proteins rich in disorder-promoting residues (DPR), cyan bars refer to Group D proteins, defined above. Note, in group D, the high occurrence of cysteine and glycine residues.

## Discussion and Conclusions

What are the determinants of protein folding has been matter of many debates in the literature until now [3, 5, 6, 16, 17]. On one side, amino acid composition is considered as very relevant as it is related to the hydrophobic effect [5, 6]. Following this view, highly hydrophilic proteins cannot form the hydrophobic core typical of globular proteins and tend to remain natively unfolded, with an internal free energy insufficient to compensate the loss of conformational entropy, necessary for the folding process [7, 8]. At variance with this view, it has been pointed out that there is no systematic evidence that amino acid composition affects thermodynamic stability of proteins [19]. Moreover, it has been observed that hydrogen-bonding, almost insensitive to amino acid composition, can however significantly contribute to the energetics of a fold [17]. Getting or not getting a stable fold, in this view, would not be predominantly determined by amino acid composition, but also influenced by the number of hydrogen-bonds that the protein can make, both internal and external, and the latter modulated by the environment (e. g. by the presence of osmolytes) [16].

To offer some information on the controversy, we analyzed 616 non ligated, non-redundant and not disordered structures from the PDB, with the aim of checking whether an unbalance of disorder promoting and order promoting residues can be related to changes in thermodynamic stability of protein folds. Our estimates of the internal free energy $e_{INT}$ are based on a simple and easily computable coarse-grained model, which allows an independent evaluation of contact energy $e_C$ and hydrogen-bond energy $e_{HB}$. We observe four groups of PDB structures with different thermal stability that constitute the basic geography of energy of the PDB. In the $e_{INT} - N$ plane, structures as stable as the reference standard of the GD model[19] are separated from labile, unstable structures and structures made hyper-stable by disulphide bridges. These groups, characterized by different thermal stability, display structural differences, easily grasped even at a simple visual inspection (see supplementary figure 3), that deserve further quantitative investigation.

The different levels of relative stability, as measured by $e_{INT}$ are strictly correlated with the ratio between contact energy and hydrogen-bond energy. Unstable structures have this ratio less than 1. Moreover, unstable proteins are characterized by an almost vanishing contact energy and by a small ratio between hydrophobic effect and de-hydration (de-mixing) energy. These energy terms are also easily computable and can be used for a rough localization of protein structures into the different energy regions we have proposed.

In spite of the fact that the structures of OPR and DPR proteins are well separated in the charge-hydrophobicity plane, amino acid composition alone does not determine thermal stability, though it has, particularly in short but not very short structures ($40 < N < 200$), some relevance. Enrichment in disorder-promoting residues is generally sufficient (though not necessary, as shown by the presence of OPR structures in group B together with DPR structures with the same $e_{INT}$) to make a structure less stable than the standard and can subtly modulate the ratios $e_C/e_{HB}$. The separation in the contact energy of hydrophobic effect and de-hydration (de-mixing) has shown that contact energy is not synonymous with a stabilizing hydrophobic effect, due to the mere internalization of residues and formation of contacts, but a destabilizing de-mixing term should be taken into account in the balance of contact forming interactions and can have a potential role in understanding relative stability of protein structures.

Interestingly, long (>200 residues) structures with $e_{INT}$ above -4.2 kJ/mol/res (group B) are not observed in the PDB. This observation deserves some remarks. It is tempting to assume that a very high (positive) $\Delta e_{HYD}$ is responsible for the absence of long structures due to the high cost of accommodating residues within a highly packed globular fold. Long OPR proteins could partly compensate the high $\Delta e_{HYD}$ through hydrophobic effect, due to their composition rich in order-promoting (hydrophobic) residues and get a stable fold. However, the relatively high $\Delta e_{HYD}$ term could inhibit crystallization of these proteins that are also hard to solve via NMR, because of chain length. In long DPR proteins, on the other hand, there should be an insufficient $\Delta e_{HYD}$ compensation, since the hydrophobic effect is systematically lower than in OPR proteins. In these proteins the de-mixing term is so high as to make the structure even more unstable than the few cases collected here in group A and should correspond to intrinsically disordered proteins. Further studies however are necessary to confirm this scenario.

As a consequence of these remarks it is also tempting to suggest a systematic screening of the crystallizability of long OPR and also of long DPR proteins. It is also important to remind that not only the mere amino acid composition, but also sequence directionality and the clustering of hydrophobic/hydrophilic residues should have a great relevance.

It is known that long proteins are hard to crystallize, but at least a posteriori, once a crystal structure has been obtained and deposited one could investigate, via our simple model, both the N dependence of $\Delta e_{HYD}$ and $e_{HP}$ consistently with the crystallizability of the structure. Further studies are necessary to clarify this theme. In the end, making public data on how hard is crystallization in different families of homologous sequences of different lengths (structural consortia must have those data), would be of great importance for structural genomics.

With this paper we add a contribution to the study of the relationship between thermodynamic stability, amino acid composition and length of protein structures. A similar contribution to this field is done by an interesting paper by Bastolla and Demetrius [28], though our parameterization of the internal free energy is different from theirs. We made an attempt at specifying the contribution due to the hydrophobic effect, the hydrogen-bonding and the de-hydration energy, while we did not stress the distinction between native and non-native contacts, very relevant in their analysis. From both papers it seems to emerge that long proteins are more prone either to unfold or to misfold than short ones, but the mechanism we suggest to explain this instability is different. In our view the instability of long proteins is not directly related to the number and the nature of contacts (though we do not see any contradiction with [28]), but is associated to the specific free energy cost of accommodating residues within a packed structure, requiring a high de-hydration energy that can be compensated only by OPR hydrophobic proteins, that get standard stability, and not by DPR hydrophilic ones, that, possibly are natively unfolded.

In this paper we have assumed the GD model as a reference. In this model the free energy of folding depends only on protein length, i.e. it is mere a constant multiplied by N. However, introducing a separation of the free energy in terms of contact energy and hydrogen-bonding and distinguishing hydrophobic effect from a de-mixing energy shed some light on the geography of relative thermal stabilities observed in the PDB.

## Materials and Methods

### Protein structures

The parameters of the energy model used in this paper were adjusted in accord with the database of thermal properties compiled by Robertson and Murphy [18]. It contains 67 proteins with known structure and both enthalpy and entropy of unfolding, extrapolated from calorimetry experiments. This database is assumed as a standard both in the GD model and in this work.

We used the model to gauge the energy content of selected structures from the Protein Data Bank [20]. The dataset was intended to collect typical folded proteins not bound to substrates, non-homologous and without disordered (e.g. non-observed, missing) residues. To obtain the dataset, we extracted from PDBSelect25 [29, 30, 31] a non-redundant list of protein chains with mutual sequence identity lower than 25% . From this list, we excluded chains of ligated proteins (with "COMPLEX" or "COMPLEXED" annotations in the fields of their PDB files) and with disordered residues (i. e. residues present in the SEQRES field but not in the ATOM field of PDB files). The two requirements of no "complexation" and of no disordered residues led to a final list of 616 structures.

### Strictly unanimous consensus score $S_{SU}$

We used the unanimous consensus score of folding $S_{SU}$ [21] to partition the dataset of structures selected from the PDB following the protocol given in the previous section. The structures were divided into three families with different amino acid compositions, similar to those of folded, natively unfolded and twilight zone proteins [21, 22], respectively. $S_{SU}$ combines into a consensus score the predictions of three sequence-only predictors of natively unfolded proteins of good performance: Poodle-W [32], the mean of the scores returned by VSL2 [33, 34] and the mean pair-wise energy computed following the IUPred algorithm [35]. It classifies a protein as folded if all the predictors agree in predicting it as folded; conversely, it classifies a protein as natively unfolded if all the indexes agree in predicting it as natively unfolded; proteins are classified as belonging to the twilight zone when two indexes disagree. As we can see from figure 4, $S_{SU}$ effectively divides the dataset into three families of structures: i) proteins predicted as folded are rich in order-promoting residues (OPR); ii) proteins predicted as unfolded are rich in disorder-promoting residues (DPR); iii) proteins belonging to the twilight zone (TZ) contain a balanced mixture of order-promoting and disorder-promoting residues [21]. Since disorder-promoting residues are more hydrophilic, DPR structures are, as a group, more hydrophilic than OPR ones, as shown in supplementary figure 1 (see in particular the inset in this figure). The normalized hydrophobicity was computed with the scale proposed by Kyte and Doolittle [36] .

**Internal free energy**

Thermodynamic stability of a protein is rooted in the balance between internal free energy and conformational entropy upon folding. A decrease of internal energy stabilizes, while a loss in conformational entropy contributes with a positive term and tends to destabilize the structure. The free energy of folding can be expressed as:

$$\Delta G_{FOLD} = E_{INT} - T \Delta S_{CONF} \qquad (1)$$

The free internal energy $E_{INT}$ contains an entropic term (see eq. (10) below) and can be separated into the sum of contact and hydrogen-bond energies:

$$E_{INT} = E_C + E_{HB} \qquad (2)$$

Contact energy is estimated through a form of the Miyazawa-Jernigan potential [24], representing the energy that residues must spend to break contacts with the solvent and make contacts among them. Essentially it embodies a model of the hydrophobic effect and a mixture term that follows Hildebrandt's theory [26]. It is important to note that, since the hydrophobic effect is both enthalpic and entropic, the contact energy is a free energy. Contact energy was estimated, over a given structure by the expression:

$$E_C = \sum_{p=1}^{N} \sum_{j=1}^{20} n_{i_p,j} \, e_{i_p,j} \qquad (3)$$

where $n_{i_p,j}$ is the number of contacts between residue of type $i$ at position $p$ with a residue of type $j$ and $e_{i_p,j}$ is the contact energy between residue of type $i$ at position $p$ with a residue of type $j$.

This form of contact energy is not suitable to take into account hydrogen-bonds, since it is not related to secondary structure elements [24]. Moreover, it does not consider interactions among neighbor residues. Hydrogen-bond energy is estimated through the model implemented in the DSSP algorithm [25].

$$E_{HB} = f \frac{q_1 \cdot q_2}{2} \sum_{p,q=1}^{N} \left[ \frac{1}{r_{ON}} + \frac{1}{r_{CH}} - \frac{1}{r_{OH}} - \frac{1}{r_{CH}} \right]_{p,q} \qquad (4)$$

where O, N, C and H label backbone atoms of the donor p-th and of the acceptor q-th residues, $q_1 = 0.42e$ and $q_2 = 0.20e$, are the charges attributed to the acceptor and donor groups.

Internal free energy, contact energy and hydrogen-bonds energy depends on the number N of amino acids within the protein sequence. We computed the values of internal free energy per residue $e_{INT}$ for each structures in our dataset, following the parameters given below. Internal free energy per residue $e_{INT}$ is simply internal free energy $E_{INT}$ divided by the number N of residues in the protein sequence. Analogously, we computed contact energy per residue $e_C$ and hydrogen-bond energy per residue $e_{HB}$.

$e_C$ and $e_{HB}$ are in arbitrary units and should be scaled to thermal units. To this end, we consider the following identities:

$$e_C^{(REF)} = c_C \cdot e_C \qquad (5)$$

$$e_{HB}^{(REF)} = c_{HB} \cdot e_{HB} \qquad (6)$$

The constant $c_{HB}$ was tuned to give to $e_{HB}^{(REF)}$ the value of -3.2 kJ/mol, previously proposed by Baldwin [5] as a reasonable average estimate over several determinations. In particular, $c_{HB}$ was determined using formula (8) below, where is the average value of and is the number of hydrogen-bonds present in the dataset of proteins selected by Robertson and Murphy [18].

$$\left\langle e_B^{(REF)} \right\rangle = c_{HB} \cdot \left\langle e_{HB} \right\rangle = c_{HB} \cdot \frac{1}{N_{hb}} \sum_{i=1}^{N_{hb}} (e_{HB})_i = -3.2 \qquad (7)$$

here $N_{hb}$ is the number of hydrogen bonds in the proteins of the training set and then:

$$c_{HB} = \frac{-3.2}{\left\langle e_{HB} \right\rangle} = \frac{-3.2 \cdot N_{hb}}{\sum_{i=1}^{N_{hb}} (e_{HB})_i} \qquad (8)$$

With this parameterization the hydrogen-bond energy per residue results -2.24 kJ/mol/res.
To compute the scale constant for contact energy $c_C$ one writes the free energy of a protein as follows [19]:

$$\Delta G_{FOLD} = N \cdot (\Delta h_{BURIAL} + \Delta h_{TRANSFER} - T \Delta s_{TRANSFER} - T \Delta s_{CONF}) \qquad (9)$$

By comparing equation (1) and (9), the free energy of interaction of proteins is defined as:

$$E_{INT} = N \cdot (e_C + e_{HB}) = N \cdot (\Delta h_{BURIAL} + \Delta h_{TRANSFER} - T \Delta s_{TRANSFER}) \qquad (10)$$

On the other hand, from the thermal data collected by Robertson and Murphy [18], one can deduce an internal free energy $e_{INT}^{(REF)}$ of -5.14 kJ/mol/residue. Using the above estimate of $e_{HB}^{(REF)}$ = -2.2 kJ/mol/res we obtain $e_C^{(REF)}$ = -2.94 kJ/mol/res. The constant $c_C$ is computed, following a method similar to that used to set $c_{HB}$, in order to get an average contact energy per residue of -2.9 kJ/mol/res.

**Hydrophobic effect and de-hydration energy**

Hydrophobic effect energy is proportional to the amino acid surface area buried in the interior of protein structure. Therefore, we can express the hydrophobic effect energy as:

$$E_{HYD} = k_h (ASA_F - ASA_U) \qquad (11)$$

where $ASA_F$ and $ASA_U$ are the surface area exposed to the solvent in the folded and in the unfolded state, respectively. Several values of the constant $k_h$ have been proposed. Following Baldwin [5], we use a value of 10.5 kJ/mol/Å2. $ASA_F$ was evaluated, for each PDB structure in our dataset using the DSSP program [25]. $ASA_U$ was evaluated using the scale reported by Rose et al. in [37].
As stated above, it has been reported that contact energy depends on hydrophobic effect and a mixture term that follows Hildebrandt's theory. We try to estimate these contribution by splitting contact energy into two terms:

$$E_C = E_{HP} + \Delta E_{HYD} \qquad (12)$$

where $E_{HP}$ is the contribution of the hydrophobic effect and $\Delta E_{HYD}$ is the de-hydration (or de-mixing) energy.
We consider the hydrophobic effect per residue $e_{HP}$ and the de-hydration energy per residue $\Delta e_{HYD}$, obtained by dividing the hydrophobic effect $E_{HP}$ and the de-hydration energy $\Delta E_{HYD}$ for the number of residues $N$ in the protein sequence. It is important to scale $e_{HP}$ so that it is compatible with the contact energy estimated through formula (5). To this end, we note that the internal energy is the contribution of three terms (see equation (10)). The enthalpy of burial and the entropy of transfer are stabilizing, while the enthalpy of transfer is destabilizing. Therefore, mixture energy must consider mainly the enthalpy of transfer, while the stabilizing terms must be due to the hydrophobic effect and the hydrogen-bonds. Therefore, we can write:

$$e_{HP} + e_{HB} = \Delta h_{BURIAL} - T \Delta s_{TRANSFER} \qquad (13)$$

From calorimetry data, we have $\Delta h_{BURIAL} - T \Delta s_{TRANSFER}$ = *-9.68* kJ/mol/residue. Taken $e_{HB}^{(REF)}$ = -2.2 kJ/mol/residue, we obtain $e_{HP}^{(REF)}$ = -7.48 kJ/mol/residue. We therefore adjust the hydrophobic effect energy so to obtain a value per residue of -7.5 kJ/mol/residue. The mixture energy $\Delta e_{HYD}$ is evaluated through formula (12).

## Authors' contributions

AD implemented the energy function, selected the database designed the experiments produced a draft and revised the paper; AG designed the experiment, wrote the paper; KS provided access and clues to use of POODLE-W, discussed and revised the paper. All authors agreed on the final form of the manuscript.

## Acknowledgements

An illuminating exchange of correspondence with Ugo Bastolla is gratefully acknowledged.

## Supplementary Materials

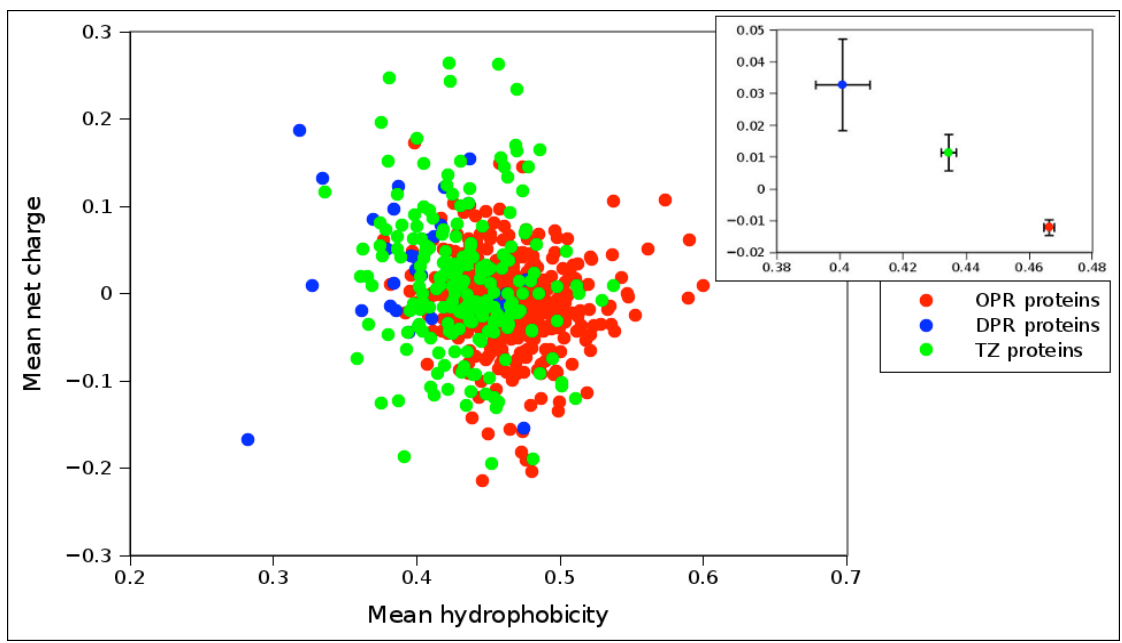

**Supplementary figure 1. Mean hydrophobicity and net charge of proteins in the PDB.**
The predictor $S_{SU}$ separates PDB representative proteins into three groups. Red dots refer to proteins rich in order-promoting residues (OPR proteins), blue dots to proteins rich in disorder-promotin residues (DPR proteins). Green dots refer to proteins that have a balanced mixture of order- and disorder-promoting residues (twilight zone proteins, TZ). In the inset the centroids of the OPR, DPR and TZ clusters are shown. Proteins in the twilight zone overlap with both OPR and DPR proteins, but their highest concentration has mean hydrophobicity in-between those of OPR and DPR proteins.

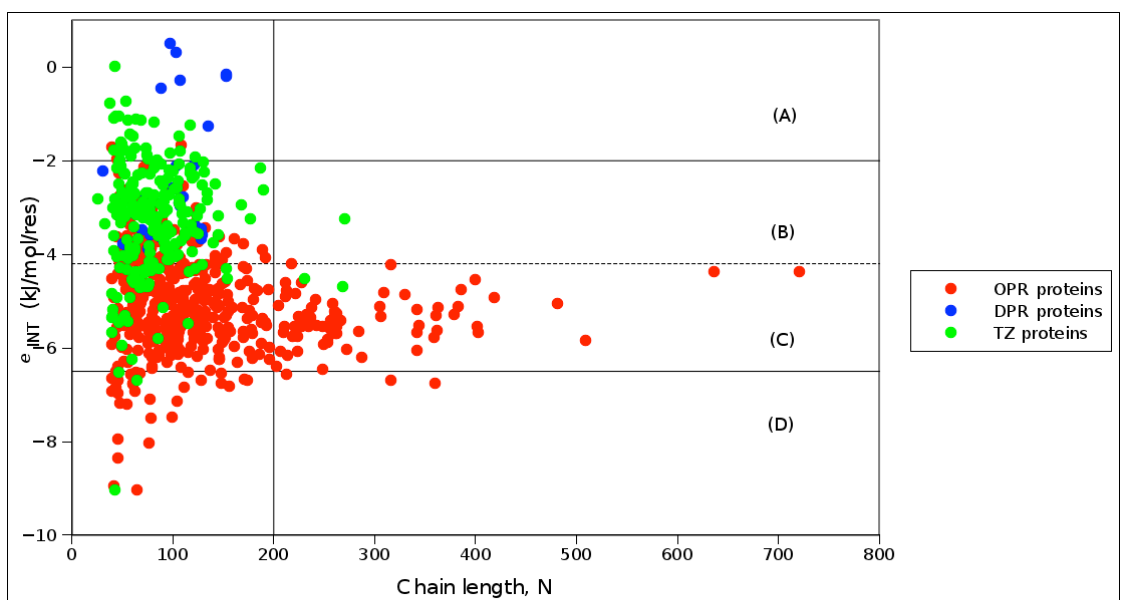

**Supplementary figure 2. TZ Proteins in the twilight belong mainly to group B**
Red dots refer to proteins rich in order-promoting residues (OPR), blue dots to proteins rich in disorder-promoting residues (DPR), green dots refer to proteins in the twilight zone. Proteins in the twilight zone have in most cases $E_{INT}$ between -2 to -4.5 kJ/mol/res (group B) or higher (group A), but very short TZ structures (peptides) are present in all the other groups.

**ENERGETICS OF PROTEINS IN THE TWILIGHT ZONE**

In the following paragraph we present estimates of various energy terms for the proteins that $S_{SU}$ unclassified as belonging to the twilight zone and that have an hydrophobicity which is intermediate between OPR and DPR proteins.

| Structural group | $<e_{INT}>$ (kJ/mol/res) | $<e_C>$ (kJ/mol/res) | $<e_{HB}>$ (kJ/mol/res) | $<e_C/e_{HB}>$ |
|---|---|---|---|---|
| A | -1.4 ± 0.1 | 0.2 ± 0.1 | -1.24 ± 0.07 | 0.2 ± 0.2 |
| B | -3.07 ± 0.05 | -1.31 ± 0.06 | -1.76 ± 0.04 | 0.86 ± 0.05 |
| C | -4.89 ± 0.09 | -3.0 ± 0.1 | -1.85 ± 0.08 | 1.8 ± 0.2 |
| D | -7.4 ± 0.6 | -5.6 ± 0.6 | -1.89 ± 0.08 | 2.9 ± 0.4 |

**Supplementary table 1. Contributions of contact energy and hydrogen-bonds to the internal free energy of PDB structures with twilight zone (TZ) amino acid composition**
As shown in the preceding supplementary figure 2 TZ structures are short and have $<e_{INT}>$ comprised in most cases between -2 to -4.5 kJ/mol/res, though very short TZ structures are present in all other groups.

| Protein groups | $<e_C>$ (kJ/mol/res) | $<e_{HP}>$ (kJ/mol/res) | $<\Delta e_{HYD}>$ (kJ/mol/res) |
|---|---|---|---|
| A | 0.2 ± 0.1 | -6.2 ± 0.4 | 6.0 ± 0.4 |
| B | -1.31 ± 0.06 | -7.8 ± 0.2 | 6.5 ± 0.2 |
| C | -3.0 ± 0.1 | -7.9 ± 0.3 | 4.9 ± 0.3 |
| D | -5.6 ± 0.6 | -6.9 ± 0.5 | 1.3 ± 1.0 |

**Supplementary table 2. Splitting contact energy of TZ PDB structures into hydrophobic effect and de-hydration energy terms**

Comparing supplementary table 1 with table 1 in the body of the paper shows that TZ proteins typically have values of $<e_{INT}>, <e_C>$ and $<e_{HB}>$ which are comparable with that of DPR proteins.

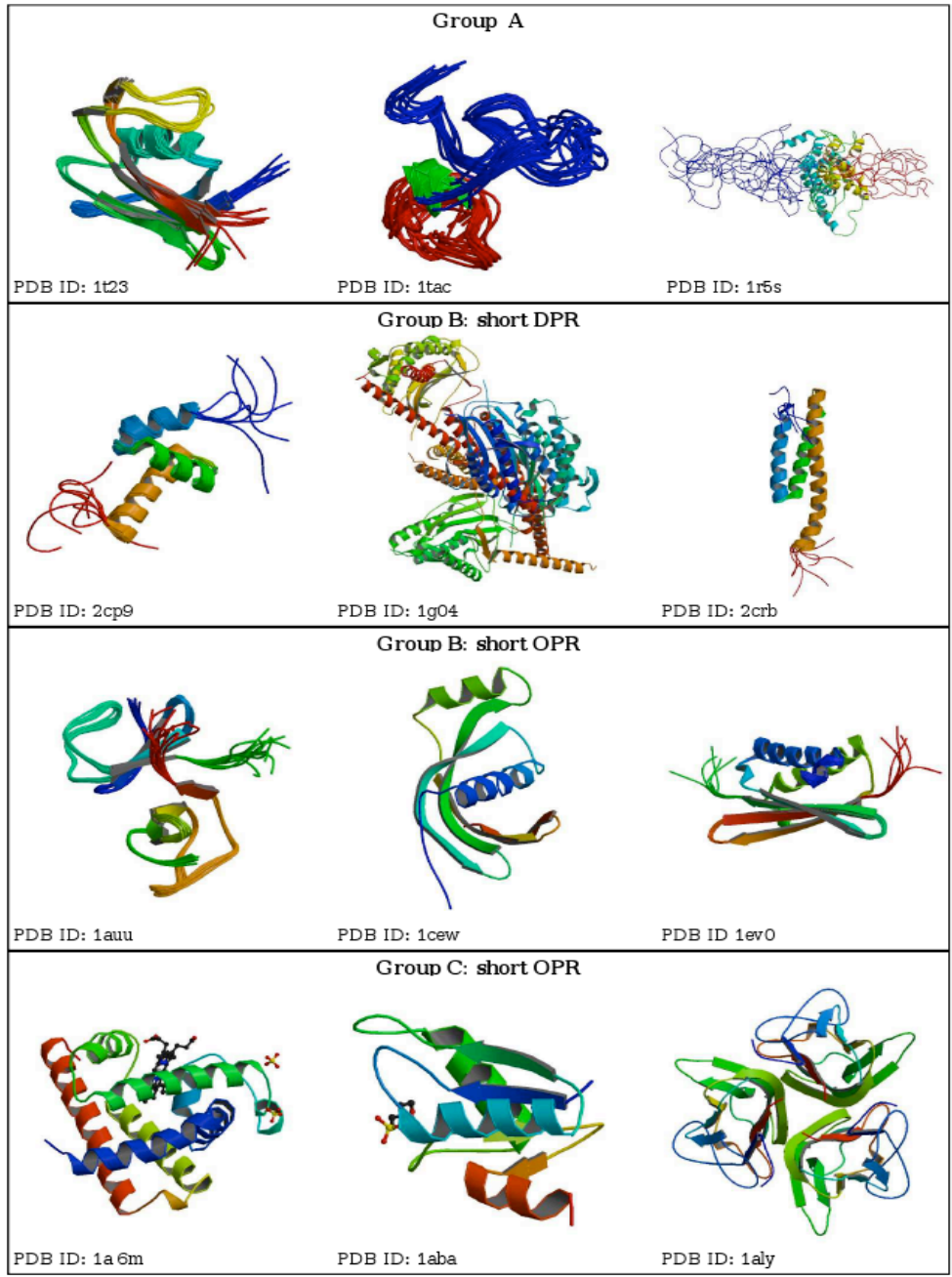

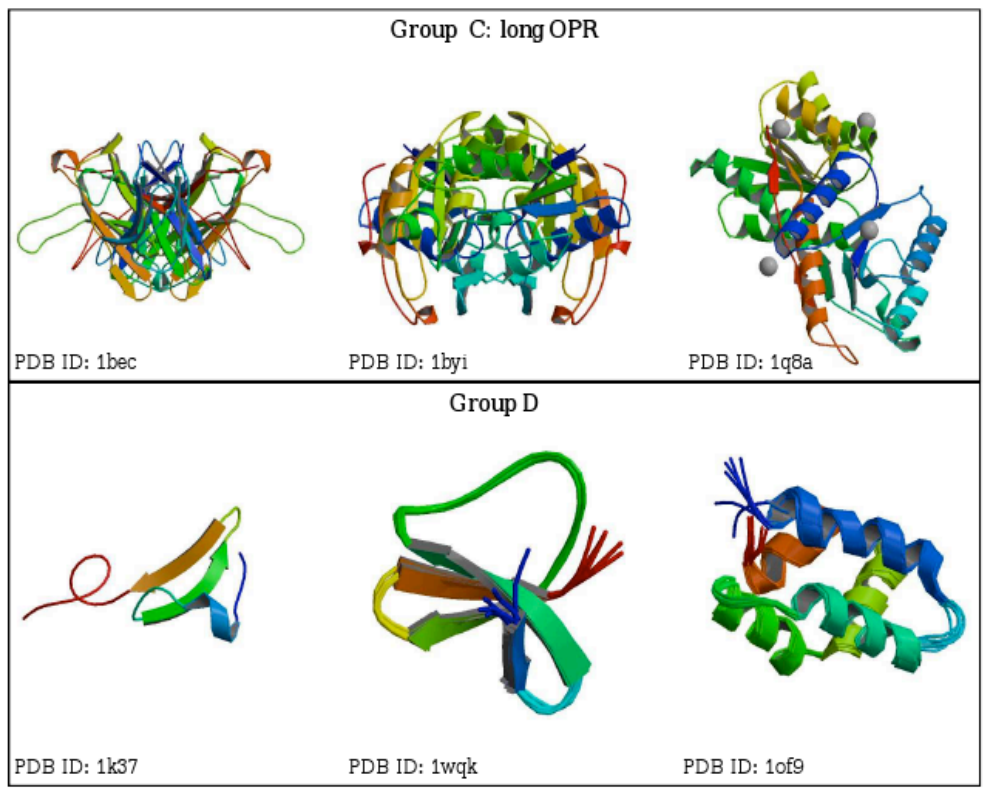

**Supplementary figure 3.** Selected structures from each group considered.